\documentstyle[11pt]{article}
\begin{document}

\title{Invariant description of solutions of hydrodynamic type systems in
hodograph space:
hydrodynamic surfaces}

\author{\Large Ferapontov E.V. \\
    Department of Mathematical Sciences \\
    Loughborough University \\
    Loughborough, Leicestershire LE11 3TU \\
    United Kingdom \\
    e-mail: \\
    {\tt E.V.Ferapontov@lboro.ac.uk}
}
\date{}
\maketitle

\newtheorem{theorem}{Theorem}
\newtheorem{proposition}{Proposition}
\newtheorem{lemma}{Lemma}

\pagestyle{plain}

\maketitle

\begin{abstract}
Hydrodynamic surfaces are solutions of  hydrodynamic type systems viewed as
non-parametrized submanifolds of the hodograph space. We propose an
invariant 
differential-geometric characterization of hydrodynamic surfaces by
expressing the curvature form of the
characteristic web in terms of the reciprocal invariants.

\bigskip

Subj. Class.: Differential Geometry, Partial Differential Equations.



\end{abstract}

\section{Introduction}

Equations of hydrodynamic type
\begin{equation}
u^i_t=v^i_j(u) u^j_x, ~~~ i, j =1, ..., n
\label{hyd}
\end{equation}
naturally arise in applications in gas dynamics, hydrodynamics, chemical
kinetics, the Whitham averaging procedure,
differential geometry and topological field theory. We refer to
\cite{Tsarev} and \cite{DN} for their geometric theory.
Any solution $u^i(x, t)$ of system (\ref{hyd}) defines a surface in the
hodograph space $u^1, ..., u^n$ parametrized by
the independent variables $x$ and $t$. Representing this surface explicitly
in the form $u^3=u^3(u^1, u^2), ..., u^n=u^n(u^1, u^2)$,
one can readily rewrite (\ref{hyd}) as a system of PDEs for $u^3, ..., u^n$
viewed as functions of
$u^1, u^2$. For $n=3$ this will be a third order
quasilinear PDE for $u^3(u^1, u^2)$,  known as the equation of a {\it
hydrodynamic surface} (by the term hydrodynamic surfaces
we mean solutions of (\ref{hyd}) viewed as non-parametrized submanifolds of
the hodograph space).
This equation was derived by Yanenko  \cite{Yanenko} and subsequently
discussed in \cite{Sidorov} in the case of
1-dimensional polytropic gas dynamics (see  Example 4 below). For
3-component systems in Riemann invariants, the derivation
of this equation is given in the Appendix. Although, technically, this
derivation does not cause any problems, the result
is usually a rather complicated PDE. This is not surprising since the choice
of, say,  $u^3$ as a function of $u^1, u^2$
introduces an
asymmetry in our approach from the very beginning. Thus, it is desirable to
have an invariant coordinate-free description of
hydrodynamic surfaces. In order to provide such a description (in what
follows we concentrate on the case $n=3$),
we first have to supply the hodograph space with certain
differential-geometric objects. These objects, known as the {\it reciprocal
invariants}, are introduced in section 3. Reciprocal
invariants induce further  geometric structure on any hydrodynamic surface
(see section 4), the characteristic 3-web being part of it.
Calculating the curvature of the characteristic 3-web in terms of the
reciprocal invariants, we obtain a simple invariant description
of hydrodynamic surfaces (Theorem 1). We apply our results to linearly
degenerate 
semi-Hamiltonian systems in Riemann invariants (section 5) and the equations
of associativity of 2-dimensional topological field theory
(section 6). In both cases the characteristic 3-web proves to be {\it
hexagonal}, thereby providing a simple coordinate-free description of
hydrodynamic surfaces.
We illustrate the concept of a hydrodynamic surface by concluding this
introduction with a list  of examples.

{\bf Example 1.} The general solution of the linear system
\begin{equation}
R^1_t=0, ~~ R^2_x=0, ~~ R^3_t=R^3_x
\label{lin}
\end{equation}
is given by
$
R^1=g^1(x), ~ R^2=g^2(t), ~ R^3=g^3(x+t)
$
where $g^i$ are arbitrary functions of their arguments. These formulae can
be rewritten as
$
x=f^1(R^1), ~ t=f^2(R^2), ~ -x-t=f^3(R^3)
$
where the functions $f^i$ are determined by $g^i$. Adding these equations,
we arrive at the hydrodynamic surfaces
\begin{equation}
f^1(R^1)+f^2(R^2)+f^3(R^3)=0
\label{linsurf}
\end{equation}
in the hodograph space $R^1, R^2, R^3$.
Notice that  surfaces (\ref{linsurf}) solve the third order PDE
\begin{equation}
\left(\ln \frac{R^3_1}{R^3_2}\right)_{12}=0,
\label{lineq}
\end{equation}
which is thus the equation of a hydrodynamic surface. Here subscripts denote
differentiation with respect to $R^1$ and $ R^2$.

{\bf Example 2.} Let us consider the system
\begin{equation}
R^1_t=(R^2+R^3)R^1_x, ~~ R^2_t=(R^1+R^3)R^2_x, ~~ R^3_t=(R^1+R^2)R^3_x
\label{weak}
\end{equation}
which, upon the introduction of the new independent variables $X$ and $ T$,
\begin{equation}
dX=\frac{dx+(R^2+R^3)dt}{(R^1-R^2)(R^1-R^3)}, ~~~
dT=\frac{dx+(R^1+R^3)dt}{(R^2-R^1)(R^2-R^3)},
\label{recip}
\end{equation}
(notice that these 1-forms are closed by virtue of (\ref{weak})),  takes the
linear form (\ref{lin}).
Transformations of this type are called {\it reciprocal}. Since any
reciprocal transformation is just a
reparametrization of solutions,  hydrodynamic surfaces of both equations
(\ref{lin}) and (\ref{weak}) are the same, defined
by the PDE (\ref{lineq}). This example is generalized in section 5 to
arbitrary three-component linearly degenerate
semi-Hamiltonian systems in Riemann invariants, which are all reciprocally
related to (\ref{lin}) and, thus, have the same
hydrodynamic surfaces.

Example 2 shows that the equation governing hydrodynamic surfaces must be
expressible in terms of
{\it reciprocal invariants},  differential-geometric objects in the
hodograph space which do not
change under reciprocal transformations.

{\bf Example 3.} The equation of associativity of 2-dimensional topological
field theory 
\begin{equation}
f_{ttt}=f_{xxt}^2-f_{xxx}f_{xtt},
\label{ass}
\end{equation}
rewritten in the variables $a=f_{xxx}, \ b=f_{xxt}, \ c=f_{xtt}$, takes the
form of a three-component system
of hydrodynamic type
\begin{equation}
a_t=b_x, ~~~ b_t=c_x, ~~~ c_t=(b^2-ac)_x.
\label{ass1}
\end{equation}
Here we list the hydrodynamic surfaces (in the hodograph space $a, b, c$)
corresponding 
to some  explicit solutions of equation (\ref{ass}) as found in
\cite{Dubrovin}.

\vspace{3ex}
\begin{tabular}{|l|l|}                                         \hline
                                          &                     \\ [-2ex]
solution of  (\ref{ass})                                    & hydrodynamic
surface in the hodograph space $a, b, c$ \\
                                                                 [1ex]
\hline
                                          &                      \\ [-2ex]
$f=\frac{x^2t^2}{4}+\frac{t^5}{60}$       & plane $a=0$ \\
                                                                 [2ex]
\hline
                                          &                      \\ [-2ex]
$f=xe^t-\frac{x^2}{24}$                   & plane $b=0$ \\
                                                                 [2ex]
\hline
                                          &                      \\ [-2ex]
$f=\frac{x^2e^t}{4}+\frac{e^{2t}}{32}-\frac{x^4}{48}$        &  quadric
$2ab+c=0$ \\
                                                                 [2ex]
\hline
                                          &                      \\ [-2ex]
$f=\frac{t^2\ln x}{2}+\frac{3t^2}{4}$              & quadric $b^2-ac=0$
\\ 
                                                                 [2ex]
\hline
                                          &                      \\ [-2ex]
$f=\frac{x^3t}{6}+\frac{x^2t^3}{6}+\frac{t^7}{210}$ &  Cayley cubic
$c-2ab+2a^3=0$ \\
                                                                 [2ex]
\hline
                                          &                      \\ [-2ex]

$f=\frac{x^3t^2}{6}+\frac{x^2t^5}{20}+\frac{t^{11}}{3960}$ & quartic
$4ac-4a^2(b-a^2/2)=(b-a^2/2)^2$ \\
                                                                   [1ex]
\hline
\end{tabular}

\vspace{3ex}

\noindent It would be interesting to classify the solutions of (\ref{ass})
whose hydrodynamic surfaces are algebraic.
A differential-geometric characterization of hydrodynamic surfaces of the
system (\ref{ass1}) is given in section 6.

{\bf Example 4.} In Lagrangian coordinates, the equations of one-dimensional
polytropic gas dynamics take the form
$$
u_t+p_q=0, ~~ \psi p^{-\kappa}p_t+u_q=0, ~~  \psi_t=0, ~~~~~
\kappa=\frac{\gamma+1}{\gamma}.
$$
The equation governing hydrodynamic surfaces $\psi=\psi(p, u)$ was derived
in \cite{Sidorov}. 
After being integrated once, it reduces to the second order quasilinear PDE
$$
(\psi p^{-\kappa}\psi _u)_u-\psi_{pp}=f(\psi)(\psi_p^2-\psi
p^{-\kappa}\psi_u^2)
$$
where $f$ is an arbitrary function of $\psi$.

\section{Exterior representation of hydrodynamic type systems}

Let $v^i$ be the eigenvalues of the matrix $v^i_j$,
called the {\it characteristic velocities} of system (\ref{hyd}), which we
assume to be real and pairwise distinct.
Let $l^i=(l^i_1(u), ..., l^i_n(u))$ be the corresponding left eigenvectors,
$l^i_jv^j_k=v ^i l^i_k.$
With the eigenforms $\omega ^i=l^i_jdu^j$,
the system (\ref{hyd}) is readily rewritten  in the  exterior form,
\begin{equation}
\omega ^i\wedge (dx+v^i dt)=0, ~~~~ i=1, ..., n.
\label{exterior}
\end{equation}
Differentiation of $\omega ^i$ and $v ^i$ gives the {\it structure
equations} 
\begin{equation}
d\omega ^i=-c^i_{jk}\omega ^j\wedge \omega^k, ~~~ (c^i_{jk}=-c^i_{kj}), ~~~
dv ^i=v ^i_j\omega ^j,
\label{structure}
\end{equation}   
containing all the necessary information about the system under study. For
diagonalizable systems we have $c^i_{jk}=0$, so that
$\omega^i=dR^i$, where the variables $R^i$ are called the {\it Riemann
invariants} of system (\ref{hyd}).
In Riemann invariants, equations (\ref{exterior}) take a diagonal form,
\begin{equation}
R^i_t=v^i(R) R^i_x, ~~~ i=1, ..., n.
\label{Riemann}
\end{equation}
In this case $v^i_j=\partial v^i/\partial R^j$.

\section{Reciprocal transformations and reciprocal invariants}

We recall the necessary information about  reciprocal transformations of
hydrodynamic type systems.
Let $B(u)dx+A(u)dt$ and $N(u)dx+M(u)dt$ be two  conservation laws of system
(\ref{hyd}), understood as
the one-forms which are closed by virtue of (\ref{hyd}).
 In the new independent variables $X,T$ defined by
\begin{equation}
dX=B(u)dx+A(u)dt,~~~dT=N(u)dx+M(u)dt,
\label{reciprocal}
\end{equation}
the system (\ref{hyd}) takes the form
\begin{equation}
u^i_T=V^i_j(u)  u^j_X,
\label{3hr}
\end{equation}
where $V=(Bv-AE)(ME-Nv)^{-1},~E=id.$
The new characteristic velocities $V^i$ are
\begin{equation}
V^i=\frac{v^iB-A}{M-v^iN},
\label{newEigen}
\end{equation}
while the eigenforms $\omega^i$ remain the same. Transformations
(\ref{reciprocal}) are called {\it reciprocal}. We refer to \cite{Rogers}
for the discussion of their
applications in gas dynamics, hydrodynamics and soliton theory. In
\cite{recip1} and \cite{recip2} the author
introduced a number of objects in the hodograph space which prove to be
reciprocally invariant. In the 3-component case these are

\noindent 1. Three 2-dimensional characteristic distributions
\begin{equation}
\omega^i=0, ~~~ i=1, 2, 3.
\label{dist}
\end{equation}
\noindent Notice that the eigenforms $\omega^i$ are only defined
\noindent up to a nonzero multiple ($\omega^i \to p^i \omega^i$), so that
only these distributions  make an invariant sense.
\noindent It is natural to call these distributions {\it characteristic}
since, by virtue of (\ref{exterior}),
\noindent characteristic directions are the intersections of a tangent plane
to a hydrodynamic surface with the
\noindent distributions (\ref{dist}).

\noindent 2. The 1-forms
\begin{equation}
\begin{array}{c}
\Omega^1=\frac{v^1_1(v^2-v^3)}{(v^1-v^2)(v^1-v^3)}\ \omega^1, ~~~
\Omega^2=\frac{v^2_2(v^3-v^1)}{(v^2-v^1)(v^2-v^3)}\ \omega^2, ~~~
\Omega^3=\frac{v^3_3(v^1-v^2)}{(v^3-v^1)(v^3-v^2)}\ \omega^3.
\end{array}
\label{Omi}
\end{equation}
\noindent In the case $v^i_i\ne 0$ these 1-forms contain the information
about the distributions (\ref{dist}).
\noindent We really need (\ref{dist}) only when some of the $v^i_i$ are
zero. 

\noindent 3. The differential $d\Omega$ of the 1-form
\begin{equation}
\begin{array}{c}
\Omega=\left(\frac{v^2_1-\frac{1}{2}v^1_1}{v^1-v^2}+
\frac{v^3_1-\frac{1}{2}v^1_1}{v^1-v^3} \right)\ \omega^1+
\left(\frac{v^1_2-\frac{1}{2}v^2_2}{v^2-v^1}+
\frac{v^3_2-\frac{1}{2}v^2_2}{v^2-v^3} \right)\ \omega^2+
\left(\frac{v^1_3-\frac{1}{2}v^3_3}{v^3-v^1}+
\frac{v^2_3-\frac{1}{2}v^3_3}{v^3-v^2} \right)\ \omega^3
\end{array}
\label{Om}
\end{equation}
(notice that $\Omega$ itself is not reciprocally invariant).

{\bf Remark.} To prove the reciprocal invariance of these objects, it is
sufficient to consider reciprocal transformations
of the two simpler types, namely

\noindent (a) The interchange of the independent variables $x$ and $t$
($A=N=1, \ B=M=0$ in (\ref{reciprocal}), implying
$V^i=1/v^i$).

\noindent (b) Transformations which preserve the variable $t$ ($N=0, \ M=1$
implying $V^i=v^iB-A$).

\noindent In both cases the invariance of  (\ref{Omi}) and (\ref{Om}) is a
result of a simple calculation.
In the case (b) one has to use the identities $A_i=v^iB_i$ (no summation!)
which characterize conservation laws of
hydrodynamic type systems.
Here $A_i$ and $B_i$ are defined by the expansions $dA=A_i\omega^i, \ dB=B_i
\omega^i$. Since an arbitrary reciprocal transformation is a superposition
of  transformations (a) and (b), the statement follows.

\medskip
Although the set of reciprocal invariants (\ref{Omi}) and (\ref{Om}) is
complete \cite{recip1}, \cite{recip2},
further invariants can be easily constructed. For instance, the
pseudo-Riemannian metric
$$
\begin{array}{c}
\Omega^1 \Omega^2+\Omega^1 \Omega^3+\Omega^2 \Omega^3=

\frac{v^1_1v^2_2}{(v^1-v^2)^2} \omega^1 \omega^2+
\frac{v^1_1v^3_3}{(v^1-v^3)^2} \omega^1 \omega^3+
\frac{v^2_2v^3_3}{(v^2-v^3)^2} \omega^2 \omega^3
\end{array}
$$
also is reciprocally invariant.
The Lie-geometric interpretation of  reciprocal invariants  was proposed in
\cite{Fer00}.
Notice that for {\it linearly degenerate} systems (that is, for systems with
$v^i_i=0$ for any $i$),
some of the reciprocal invariants vanish. The only that survive are the
2-form $d\Omega$ and the
characteristic distributions
 (which are not involutive unless the system possesses Riemann invariants).
Reciprocal transformations are known to preserve the linear degeneracy.

\section{Geometry of hydrodynamic surfaces}

The following objects are naturally induced on hydrodynamic surfaces of
3-component systems of hydrodynamic type.

\noindent {\bf The characteristic 3-web} is a collection of three
1-parameter families of curves defined by the equations
$\omega^i=0$. Geometrically, characteristic directions are intersections of
the tangent plane of a hydrodynamic surface with
the 2-dimensional characteristic distributions $\omega^i=0$ in the hodograph
space. For a 3-web on a surface,
let $\phi^1$ and $\phi^2$ be the 1-forms such that the curves of the first,
second and third families are defined by the
equations $\phi^1=0, \ \phi^2=0$ and $\phi^1=\phi^2$, respectively
(actually, $\phi^1$ and $\phi^2$ are the properly normalized
$\omega^1$ and $\omega^2$; their choice is obviously non-unique since they
can be multiplied by a common factor).
The connection form $\phi$ of a 3-web is uniquely determined by the exterior
equations
$$
d \phi ^1=\phi \wedge \phi ^1, ~~~ d \phi ^2=\phi \wedge \phi ^2.
$$
Finally, the curvature 2-form $C$ equals $d\phi$. It has an invariant
meaning
and does not depend on the particular normalization of $\phi^1$ and
$\phi^2$. For instance, in the case of 3-component systems
(\ref{Riemann}) in 
Riemann invariants  we have $\omega^i=dR^i$ so that the characteristics are
defined by
$dR^1=0, \ dR^2=0$ and $dR^3=R^3_1dR^1+R^3_2dR^2=0$ (the hydrodynamic
surface is parametrized in the form $R^3(R^1, R^2)$).
Clearly, one can choose $\phi^1=R^3_1dR^1$ and $\phi^2=-R^3_2dR^2$, implying
that the connection form is
$$
\phi=(\ln R^3_1)_2\ dR^2+(\ln R^3_2)_1\ dR^1.
$$ 
The corresponding curvature 2-form is
$$
C=d\phi=(\ln R^3_1-\ln R^3_2)_{12}\ dR^1\wedge dR^2.
$$
We will use this expression when deriving the equation of
hydrodynamic surfaces in the Appendix. Recall that the zero curvature webs
are called hexagonal \cite{Blaschke}.

\bigskip

\noindent {\bf The form  $d\Omega$} is just the restriction of the
reciprocal invariant $d\Omega$ to a hydrodynamic surface.

\bigskip

\noindent {\bf The forms $*\Omega^i$} are defined as follows. Take, say, the
reciprocally invariant form $\Omega^1$
and restrict it to a hydrodynamic surface. On the same hydrodynamic surface,
choose
the  metric $2dR^2dR^3$ with  the volume form
$dR^2\wedge dR^3$ (notice the important order of  indices $2$ and $3$ in the
expressions for 
$\Omega^1$ and the volume form).
Construct the vector which is dual to the 1-form $\Omega^1$ with respect to
the metric choosen. Finally, evaluate the
volume form on this vector. The result will be a 1-form which is usually
denoted by $*\Omega^1$. Notice that in
two dimensions the $*$-operator is conformally invariant, so that only the
conformal class of the metric $2dR^2dR^3$ matters.
The ambiguity in the choice of the order of indices $2$ and $3$ in the
expression for $\Omega^1$, combined
with the ambiguity in choosing the sign (orientation) of the volume form,
gives a well-defined answer for
$*\Omega^1$. In general, for the 1-form
$$
\Omega^i=\frac{v^i_i(v^j-v^k)}{(v^i-v^j)(v^i-v^k)}\ dR^i,
$$
the 1-form $*\Omega^i$ is defined by choosing the metric $2dR^jdR^k$ with
the volume form $dR^j\wedge dR^k$. The $*$-operator
is invariant and can be applied in any convenient coordinate system leading
to one and the same result. The
relevant computation is shown below in the proof of Theorem 1.

\bigskip

\noindent Now we are in order to formulate the main result of this paper.

\begin{theorem}
The curvature 2-form $C$ of the characterstic 3-web is given by the formula
\begin{equation}
C=-d\Omega-\frac{1}{2}d(*\Omega^1+*\Omega^2+*\Omega^3).
\label{C}
\end{equation}
 Written down in any suitable coordinate system
in the hodograph space, equation (\ref{C}) reduces to a third order PDE for
hydrodynamic surfaces.
\end{theorem}

\medskip

\centerline{\bf Proof:}

\medskip

\noindent Notice that equation (\ref{C}) is manifestly coordinate-free.
Thus, it sufficies to establish (\ref{C}) in any
local parametrization of a hydrodynamic surface. We will work directly in
the coordinates $x, t$. For simplicity, we assume
that the system in question possesses Riemann invariants. This assumption is
not important and the general proof is literally the same.
Introducing the 
1-forms
$$
\phi ^1=(v^2-v^3)(dx+v^1dt) ~~ {\rm and} ~~ \phi ^2=(v^1-v^3)(dx+v^2dt),
$$
we readily see that the characteristics of the first, second and third
families 
are defined by the equations
$\phi^1=0, \ \phi ^2=0$ and $\phi ^1=\phi ^2$, respectively. The connection
form $\phi$
of the 3-web is uniquely determined by the exterior equations
\begin{equation}
d \phi ^1=\phi \wedge \phi ^1, ~~~ d \phi ^2=\phi \wedge \phi ^2.
\label{c}
\end{equation}
 With
$\phi=adx+bdt$, the  equation $(\ref{c})_1$ gives
$$
b-av^1=-v^1_1R^1_x+\left((v^2-v^1)\frac{v^2_2-v^3_2}{v^2-v^3} -v^1_2
\right)R^2_x+
\left((v^3-v^1)\frac{v^2_3-v^3_3}{v^2-v^3} -v^1_3 \right)R^3_x.
$$
Similarly, the  equation $(\ref{c})_2$ implies
$$
b-av^2=-v^2_2 R^2_x+\left((v^1-v^2)\frac{v^1_1-v^3_1}{v^1-v^3} -v^2_1
\right) R^1_x+
\left((v^3-v^2)\frac{v^1_3-v^3_3}{v^1-v^3} -v^2_3 \right) R^3_x.
$$
Solving for $a$ and $b$, we obtain
$$
\begin{array}{c}
a=\left(\frac{v^2_1-v^1_1}{v^2-v^1}+\frac{v^3_1-v^1_1}{v^3-v^1}\right)
R^1_x+
\left(\frac{v^1_2-v^2_2}{v^1-v^2}+\frac{v^3_2-v^2_2}{v^3-v^2}\right) R^2_x+
\left(\frac{v^1_3-v^3_3}{v^1-v^3}+\frac{v^2_3-v^3_3}{v^2-v^3}\right) R^3_x,
\\
\ \\
b=\left( \frac{v^1v^2_1-v^2v^1_1}{v^2-v^1} + v^1
\frac{v^3_1-v^1_1}{v^3-v^1} \right) R^1_x+
\left( \frac{v^2v^1_2-v^1v^2_2}{v^1-v^2} + v^2
\frac{v^3_2-v^2_2}{v^3-v^2} \right) R^2_x + \\
\ \\
\left ( v^3 \frac{v^1_3-v^3_3}{v^1-v^3} + v^3
\frac{v^2_3-v^3_3}{v^2-v^3} - v^3_3 \right ) R^3_x.
\end{array}
$$
Using the identities $dR^i=R^i_x(dx+v^idt)$, it is now a direct algebraic
calculation to verify that
\begin{equation}
\begin{array}{c}
adx+bdt=-\Omega + \frac{1}{2} v^1_1 \left(
\frac{dR^1}{v^1-v^2}+\frac{dR^1}{v^1-v^3}-2R^1_x dt \right) + \\
\ \\
\frac{1}{2} v^2_2 \left( \frac{dR^2}{v^2-v^1}+ \frac{dR^2}{v^2-v^3}-2R^2_x
dt
\right) +
\frac{1}{2} v^3_3 \left( \frac{dR^3}{v^3-v^1} + \frac{dR^3}{v^3-v^2}-2R^3_x
dt
\right) ,
\end{array}
\label{0}
\end{equation}
where the last three terms are nothing but $-\frac{1}{2}(*\Omega^1), \
-\frac{1}{2}(*\Omega^2)$ and $ -\frac{1}{2}(*\Omega^3)$,
respectively. Indeed, let us calculate $*\Omega^1$. In coordinates $x, t$
the 1-form 
$\Omega^1=v^1_1(v^2-v^3)dR^1/(v^1-v^2)(v^1-v^3)$  has components
\begin{equation}
(\mu R^1_x, \ \  \mu v^1 R^1_x)
\label{1}
\end{equation}
where $\mu=v^1_1(v^2-v^3)/(v^1-v^2)(v^1-v^3)$. The metric used to define the
$*$-operator is
$2dR^2dR^3=2R^2_xR^3_x(dx^2+(v^2+v^3)dxdt+v^2v^3dt^2)$ or, in matrix form,
\begin{equation}
R^2_xR^3_x
\left(
\begin{array}{cc}
2&v^2+v^3\\
v^2+v^3&2v^2v^3
\end{array}
\right),
\label{2}
\end{equation}
with the inverse
\begin{equation}
\frac{1}{R^2_xR^3_x(v^3-v^2)^2}
\left(
\begin{array}{cc}
-2v^2v^3&v^2+v^3\\
v^2+v^3&-2
\end{array}
\right).
\label{3}
\end{equation}
The corresponding  volume 2-form is
\begin{equation}
dR^2\wedge dR^3=R^2_xR^3_x(v^3-v^2)dx\wedge dt.
\label{4}
\end{equation}
Multiplying (\ref{1}) by (\ref{3}), we obtain the vector (the dual of
$\Omega^1$)
\begin{equation}
\frac{\mu R^1_x}{R^2_xR^3_x(v^3-v^2)^2}(v^1(v^2+v^3)-2v^2v^3, \ \
v^2+v^3-2v^1).
\label{5}
\end{equation}
Finally, evaluating the volume 2-form (\ref{4}) on the vector (\ref{5}), we
obtain the 1-form 
$$
*\Omega^1=\frac{\mu
R^1_x}{v^3-v^2}\left((v^1(v^2+v^3)-2v^2v^3)dt-(v^2+v^3-2v^1)dx\right),
$$
which, after a simple rearrangement of terms, can be rewritten as
$$
*\Omega^1=v^1_1\left(2R^1_x dt-\frac{dR^1}{v^1-v^2}-\frac{dR^1}{v^1-v^3}
\right).
$$
Comparison with (\ref{0}) and the exterior differentiation complete the
proof.

In the nondiagonalizable case the proof is essentially the same. The only
difference is that we have to write
$\omega^i=p^i(dx+v^idt)$ instead of $dR^i=R^i_x(dx+v^idt)$ and to replace
$R^i_x$ by $p^i$ in all places where they appear.
Another proof (which is more constructive, although only applies to systems
in Riemann invariants), is given in the Appendix.

\section{Linearly degenerate semi-Hamiltonian systems in Riemann invariants}

A system of hydrodynamic type in Riemann invariants,
$$
R^i_t=v^i(R) R^i_x,
$$
is called {\it linearly degenerate} if $v^i_i=0$ for any $i$. It is called
{\it semi-Hamiltonian} if
$$
\left(\frac{v^i_j}{v^j-v^i}\right)_k=\left(\frac{v^i_k}{v^k-v^i}\right)_j
$$
for any $i\ne j\ne k$. The last condition is equivalent to the existence of
an infinity 
of conservation laws and hydrodynamic symmetries and the integrability of
the system under study by the
{\it generalized hodograph transform}  \cite{Tsarev}. One can readily verify
that the reciprocal invariants
$d\Omega$ and $\Omega^1, \Omega^2, \Omega^3$ of  three-component linearly
degenerate semi-Hamiltonian systems
(the example of which is (\ref{weak}))
are zero. All such systems can be linearised by a reciprocal transformation,
and  hydrodynamic surfaces thereof are
governed by one and the same PDE (\ref{lineq}). Geometrically, hydrodynamic
surfaces are uniquely characterized as the surfaces
in the hodograph space $R^1, R^2, R^3$ on which the 3-web, cut  by the
coordinate planes $R^i=const$, is hexagonal.
We refer to \cite{Fer89} and \cite{Fer90} for a further discussion of the
geometry of characteristic webs on solutions of
linearly degenerate semi-Hamiltonian systems.

\section{Equations of associativity}

It was demonstrated in \cite{Fer96} that  system (\ref{ass1}) can be
transformed by a reciprocal transformation to a system
with constant charactersitic velocities. Since for systems with constant
characteristic velocities the objects $\Omega^i, \Omega ^2, \Omega^3$ and
$\Omega$
are automatically zero, they are zero for  system (\ref{ass1}) as well (in
view of their reciprocal invariance). Thus,
Theorem 1 implies that hydrodynamic surfaces of (\ref{ass1})  are uniquely
characterized as surfaces on
which the characteristic 3-web (cut  by the characteristic distributions
(\ref{dist})) is hexagonal.
These distributions have a simple algebro-geometric description which we
briefly discuss below (see also \cite{Fer1}).

In the hodograph space $a, b, c$ consider the twisted cubic $\gamma$,
$$
a=-3t, \ b=-3t^2/2, \ c=-t^3.
$$
For any point $p$ in the hodograph space, there are exactly three osculating
planes of $\gamma$  containing $p$.
In each of these planes, draw a line through $p$ parallel to the tangential
direction to $\gamma$ in the point where
this plane osculates $\gamma$. Thus, one obtains three lines through each
point $p$ in the hodograph space. These lines are
the  {\it rarefaction curves} of system (\ref{ass1}). The three
2-dimensional characteristic
distributions in question are spanned by each pair thereof.

This construction can be reformulated in a projectively invariant way as
follows. Consider a twisted cubic $\gamma$ in
the projective space $P^3$ and fix the plane $\Lambda$ which osculates
$\gamma$ 
(in the construction above this was the plane at infinity). Take any other
plane $\pi $ which osculates $\gamma$
in a point $p$. The tangent line to $\gamma$ in the point $p$ cuts $\Lambda$
in the point $A(p)$ (as $p$ varies,
the collection of points $A(p)$ is a conic in $\Lambda$). Finally, consider
a pencil of lines in the plane $\pi$ with the
vertex in $A(p)$. As $p$ varies, this gives a 2-parameter family, or a
congruence of lines in $P^3$, which
is of the order 3 (that is, there are precisely 3 lines of the congruence
through a generic 
point of $P^3$). The three 2-dimensional distributions are spanned by each
pair thereof. In the case when $\Lambda$ is the
plane at infinity, this contruction reduces to that described above.
Hydrodynamic surfaces in question are the ones on
which these 2-dimensional distributions cut a hexagonal 3-web. These
considerations and the Example 3 clearly indicate that
it is of interest to classify hydrodynamic surfaces which are algebraic.

Notice that this problem makes sense for  arbitrary congruences of the order
3 in $P^3$, since any such
congruence induces a 3-web on a surface.

\section{Appendix: another proof of Theorem 1}

We will derive the equation of a hydrodynamic surface for three-component
systems in Riemann invariants assuming
$R^3=R^3(R^1, R^2)$. We use the notation $R^3_1, R^3_2$ for partial
derivatives of $R^3$
with respect to $R^1$ and $R^2$ which are now viewed as independent
variables. By virtue of (\ref{Riemann}), one has
\begin{equation}
dR^1=p^1(dx+v^1dt), ~~~ dR^2=p^2(dx+v^2dt), ~~~  dR^3=p^3(dx+v^3dt)
\label{dR}
\end{equation}
(here $p^i=R^i_x$). On the other hand,
$$
dR^3=R^3_1dR^1+R^3_2dR^2=R^3_1p^1(dx+v^1dt)+R^3_2p^2(dx+v^2dt),
$$
implying
$$
R^3_1p^1+R^3_2p^2=p^3, ~~~ R^3_1p^1v^1+R^3_2p^2v^2=p^3v^3
$$
so that
$$
p^1=\frac{v^2-v^3}{v^2-v^1}\frac{p^3}{R^3_1}, ~~~
p^2=\frac{v^1-v^3}{v^1-v^2}\frac{p^3}{R^3_2}.
$$
Substituting these expressions into the first two equations (\ref{dR}), we
obtain
$$
dx+v^1dt=\frac{v^2-v^1}{v^2-v^3}\frac{R^3_1}{p^3}dR^1, ~~~
dx+v^2dt=\frac{v^1-v^2}{v^1-v^3}\frac{R^3_2}{p^3}dR^2
$$
so that
$$
dt=\frac{R^3_1}{v^3-v^2}\frac{1}{p^3}dR^1+\frac{R^3_2}{v^3-v^1}\frac{1}{p^3}
dR^2, ~~~
dx=\frac{R^3_1}{v^2-v^3}\frac{v^2}{p^3}dR^1+\frac{R^3_2}{v^1-v^3}\frac{v^1}{
p^3}dR^2.
$$
Introducing $g$ by the formula $1/p^3=e^g(v^3-v^1)(v^3-v^2)$, we ultimately
have
\begin{equation}
\begin{array}{c}
dt=e^g(R^3_1(v^3-v^1)dR^1+R^3_2(v^3-v^2)dR^2), \\
\ \\
dx=e^g(R^3_1(v^1-v^3)v^2dR^1+R^3_2(v^2-v^3)v^1dR^2).
\label{dtdx}
\end{array}
\end{equation}
With $dg=g_1dR^1+g_2dR^2$ the differentiation of  $(\ref{dtdx})_1$ implies
$$
\begin{array}{c}
g_1R^3_2(v^3-v^2)-g_2R^3_1(v^3-v^1)+R^3_{12}(v^1-v^2)+ \\
\ \\
R^3_2(v^3_1-v^2_1)-R^3_1(v^3_2-v^1_2)+R^3_1R^3_2(v^1_3-v^2_3)=0.
\end{array}
$$
Similarly, the differentiation of  $(\ref{dtdx})_2$ gives
$$
\begin{array}{c}
g_1R^3_2(v^3-v^2)v^1-g_2R^3_1(v^3-v^1)v^2+R^3_{12}(v^1-v^2)v^3+ \\
\ \\
R^3_2(v^1(v^3_1-v^2_1)+v^1_1(v^3-v^2))-R^3_1(v^2(v^3_2-v^1_2)+v^2_2(v^3-v^1)
)+\\
\ \\
R^3_1R^3_2(v^3(v^1_3-v^2_3)+(v^1-v^2)v^3_3)=0.
\end{array}
$$
Solving these equations for $g_1$ and $g_2$, one readily obtains
$$
\begin{array}{c}
-g_1=\frac{R^3_{12}}{R^3_2}+\frac{v^1_1}{v^1-v^2}+\frac{v^3_1-v^2_1}{v^3-v^2
}+\\
\ \\
\frac{R^3_1}{R^3_2}\frac{v^2_2(v^1-v^3)}{(v^2-v^1)(v^2-v^3)}
+R^3_1(\frac{v^3_3}{v^3-v^2}+\frac{v^1_3-v^2_3}{v^1-v^2})
\end{array}
$$
and
$$
\begin{array}{c}
-g_2=\frac{R^3_{12}}{R^3_1}+\frac{v^2_2}{v^2-v^1}+\frac{v^3_2-v^1_2}{v^3-v^1
}+\\
\ \\
\frac{R^3_2}{R^3_1}\frac{v^1_1(v^2-v^3)}{(v^1-v^2)(v^1-v^3)}
+R^3_2(\frac{v^3_3}{v^3-v^1}+\frac{v^1_3-v^2_3}{v^1-v^2}),
\end{array}
$$
or, in differential form,
$$
\begin{array}{c}
-dg=\frac{R^3_{12}}{R^3_2}dR^1+\frac{R^3_{12}}{R^3_1}dR^2+\frac{v^1_1}{v^1-v
^2}dR^1+\frac{v^2_2}{v^2-v^1}dR^2+\\
\ \\
\frac{v^3_1-v^2_1}{v^3-v^2}dR^1+\frac{v^3_2-v^1_2}{v^3-v^1}dR^2+\frac{v^1_3-
v^2_3}{v^1-v^2}dR^3+\\
\ \\
\frac{R^3_1}{R^3_2}\frac{v^2_2(v^1-v^3)}{(v^2-v^1)(v^2-v^3)}dR^1+
\frac{R^3_2}{R^3_1}\frac{v^1_1(v^2-v^3)}{(v^1-v^2)(v^1-v^3)}dR^2+ \\
\ \\
R^3_1\frac{v^3_3}{v^3-v^2}dR^1+R^3_2\frac{v^3_3}{v^3-v^1}dR^2.
\end{array}
$$
It is now a direct algebraic calculation to verify that terms in the last
equation can be rearranged in such a way that
it becomes
\begin{equation}
\begin{array}{c}
-dg-d\ln 
(v^1-v^2)(v^1-v^3)(v^2-v^3)=\frac{R^3_{12}}{R^3_2}dR^1+\frac{R^3_{12}}{R^3_1
}dR^2+\Omega+ \\
\ \\
\frac{v^1_1(v^2-v^3)}{(v^1-v^2)(v^1-v^3)}\left(\frac{1}{2}dR^1+\frac{R^3_2}{
R^3_1}dR^2 \right) +
\frac{v^2_2(v^1-v^3)}{(v^2-v^1)(v^2-v^3)}\left(\frac{1}{2}dR^2+\frac{R^3_1}{
R^3_2}dR^1 \right) + \\
\ \\
\frac{1}{2}\frac{v^3_3(v^1-v^2)}{(v^3-v^1)(v^3-v^2)}\left(R^3_2dR^2-R^3_1dR^
1 \right).
\end{array}
\label{g}
\end{equation}
Now, the first term on the right,
$$
\frac{R^3_{12}}{R^3_2}dR^1+\frac{R^3_{12}}{R^3_1}dR^2,
$$
is the connection form of the characteristic three-web; its differential is
the curvature form of the web.
The form $\Omega$ is defined in section 3 (recall that its differential is
reciprocally invariant).
The last three forms on the right are nothing but
$* \Omega^1/2, \ * \Omega^2/2$ and $* \Omega^3/2$, respectively. Finally,
the differentiation of (\ref{g}) completes the proof.

Notice that this proof is constructive: once a hydrodynamic surface is
given, it can be parametrized by the
independent variables $t, x$ according to the formulae (\ref{dtdx}), where
$g$ can is given by (\ref{g}). This
parametrization is unique up to the obvious symmetries $x\to cx+a, \ t\to
ct+b$ where $a, b$ and $ c$ are constants.

\section{Acknowledgements}
This research was supported by the EPSRC grant No Gr/N30941.


\begin{thebibliography}{99}
\addcontentsline{toc}{section}{References}



\bibitem{Fer1} S.~I. Agafonov and E.~V. Ferapontov, Systems of conservation
laws from the point of view of the projective theory of congruences,
Izv. RAN, ser. mat. {\bf 60} (1996) N.6, 3-30.


\bibitem{Blaschke} W. Blaschke, Einf\"uhrung in die Geometrie der Waben,
Birkh\"auser-Verlag, Basel-Stuttgart, 1955, 108 pp.

\bibitem{Dubrovin} B.~A. Dubrovin, Geometry of 2D topological field
theories, Lecture Notes in Mathematics, V.1620, Berlin,
Springer, 120-348.

\bibitem{DN} B.~A. Dubrovin and S.~P. Novikov, Hydrodynamics of weakly
deformed
soliton lattices. Differential geometry and Hamiltonian theory, Uspekhi Mat.
Nauk
{\bf 44} (1989) N6, 29-98.


\bibitem{Fer89} E.~V. Ferapontov,
Systems of three differential equations of hydrodynamic type
with hexagonal 3-web of characteristics on the solutions,
Functs. Analiz i ego Pril., {\bf 23}, N.2 (1989) 79-80.

\bibitem{recip1} E.~V. Ferapontov, Reciprocal transformations and their
invariants,  Diff. Uravnen., {\bf 25}, N.7 (1989) 1256-1265;
(English transl. in Diff. Equations, {\bf 25}, N.7 (1989) 898-905).

\bibitem{Fer90} E.~V. Ferapontov, Integration of weakly nonlinear
semi-Hamiltonian
systems of hydrodynamic type by methods of the theory of webs,
Mat. Sbornik, 181, N.9 (1990) 1220-1235.

\bibitem{recip2} E.~V. Ferapontov, Reciprocal autotransformations and
hydrodynamic symmetries, Diff. Uravnen., {\bf 27}, N.7 (1991) 1250-1263;
(English transl. in Diff. Equations, {\bf 27}, N.7 (1991) 885-895).


\bibitem{Fer96} E.~V. Ferapontov and O.~I. Mokhov, Equations of
associativity
of two-dimensional topological field theory as integrable Hamiltonian
nondiagonalisable systems of hydrodynamic type, Funkt. Anal. and it's Appl.,
{\bf 30}, N3 (1996) 62-72.

\bibitem{Fer00} E.~V. Ferapontov, Lie sphere geometry and integrable
systems, Tohoku
Math. J. {\bf 52} (2000) 199-233.

\bibitem{Rogers} C. Rogers and W.~F. Shadwick, B\"acklund Transformations
and their Applications, Academic Press, N.Y., 1982.

\bibitem{Sidorov} A.~F. Sidorov, V.~P. Shapeev and N.~N. Yanenko, The method
of differential constraints and its
applications in gas dynamics, Nauka, Novosibirsk, 1984.

\bibitem{Tsarev} S.~P. Tsarev, The geometry of Hamiltonian systems of
hydrodynamic type. The generalized hodograph transform, Math. USSR Izv.
{\bf 37} (1991) 397-419.

\bibitem{Yanenko} N.~N. Yanenko, Reduction of  a system of quasilinear
equations to a quasilinear equation,
Usp. Mat. Nauk {\bf 10} (1955), no 3 (65), 173-178.

\end{thebibliography}
\end{document}